\documentclass[10pt,letterpaper]{article}
\usepackage{opex3}
\usepackage{graphicx}

\begin{document}

\title{Modeling of mode-locking in a laser with spatially separate gain media}

\author{R.M. Oldenbeuving$^{1,*}$, C.J. Lee$^{1}$, P.D. van Voorst$^2$, H.L. Offerhaus$^3$, K.-J. Boller$^1$}

\address{
$^1$University of Twente, Laser Physics and Nonlinear Optics
group, MESA+ Research Institute for Nanotechnology, P.O. Box 217,
7500AE, Enschede, The Netherlands\\ $^2$Sensor Sense, P.O. box
9010, 6500GL Nijmegen, The Netherlands\\ $^3$University of Twente,
Optical Sciences group, MESA+ Research Institute for
Nanotechnology, P.O. Box 217, 7500AE, Enschede, The Netherlands
}
{\vskip-.3cm \parskip0pc\hskip2.25pc \footnotesize%
 \parbox{.8\textwidth}{\begin{center}$^*$\it \textcolor{blue}{\underline{R.M.Oldenbeuving@UTwente.nl}} \rm \end{center} } \normalsize  \vskip-.2cm}
\homepage{http://lpno.tnw.utwente.nl/}


\begin{abstract}
We present a novel laser mode-locking scheme and discuss its
unusual properties and feasibility using a theoretical model. A large
set of single-frequency continuous-wave lasers oscillate by
amplification in spatially separated gain media. They are mutually
phase-locked by nonlinear feedback from a common saturable
absorber. As a result, ultra short pulses are generated. The new
scheme offers three significant benefits: the light that is
amplified in each medium is continuous wave, thereby avoiding
issues related to group velocity dispersion and nonlinear effects
that can perturb the pulse shape. The set of frequencies on which
the laser oscillates, and therefore the pulse repetition rate, is
controlled by the geometry of resonator-internal optical elements,
not by the cavity length. Finally, the bandwidth of the laser can
be controlled by switching gain modules on and off. This scheme
offers a route to mode-locked lasers with high average output
power, repetition rates that can be scaled into the THz range, and
a bandwidth that can be dynamically controlled. The approach is
particularly suited for implementation using semiconductor diode
laser arrays.
\end{abstract}



\section{Introduction}
Ultra-short laser pulses are widely used in scientific research ranging
from fundamental physics to biology, and
find use in many applications such as for material processing or in
telecommunications~\cite{NatureKlein2010,NatureKeller2003}.
The most common
method to generate ultra-short pulses is passive mode-locking
because this can yield particularly short pulse durations in the
femtosecond range. However, current mode-locking methods place
restrictions on the combinations of repetition rates, output
powers, and pulse durations that are feasible, because the pulse
repetition rate scales inversely with the length of the laser cavity~\cite{Pfeiffer1993}.

The connection between repetition rate and cavity length implies
that high repetition rate and high average power are incompatible
using current approaches. To illustrate this, let us consider a
very short cavity length of a few hundred $\mu m$, such as can be
realized in semiconductor lasers. Indeed, very high repetition
rates of hundreds of GHz can be obtained~\cite{Keller2006}.
However, the short cavity length limits the roundtrip gain and
average power, typically to the mW-range with such semiconductor
lasers. Extending the cavity length allows for higher average
output power, but also lowers the repetition rate and thereby
introduces other limitations. A first restriction, noticeably in
semiconductor lasers, comes from the limited lifetime
(nanoseconds) of the upper state, such as given by spontaneous
emission. At repetition rates of a few GHz and below, amplified
spontaneous emission occurs between pulses~\cite{Robertson2000},
which can deplete the inversion and reduce the coherence of the
output. A second restriction is distortion of the pulse shape by
dispersion, gain saturation, and other nonlinear effects. Other
mode-locked lasers, for instance Ti:Sappire oscillators, have an
average output power limited to a few Watts by such effects.

The described incompatibility limits applications where nonlinear processes need to be driven at the highest repetition
frequency. For example, photomagnetic
switching~\cite{Hansteen2005} requires, 50~GW/cm$^2$ (over 100 fs)
to initiate a switching operation. A 1~THz pulse repetition rate
laser, focused to a diffraction limited spot, therefore, requires
an average power of approximately 10~W, which is well beyond the
range of the capabilities of current mode-locked diode
lasers~\cite{Aschwanden2005}. Another example is all optical
switching~\cite{Harding2007} with similar
requirements.

We investigate a novel approach to mode-locking in which multiple
single-frequency lasers are phase-locked via a common non-linear
optical element. Our approach may be applied to all types of
lasers, however, for clarity and relevance, we concentrate on
diode lasers. To illustrate the scheme and investigate its
feasibility with current technology, we discuss a specific setup,
in the form of an external cavity diode laser using bulk optical
elements as shown in Fig.~\ref{FigExperimentalSetup}, although one
could also think of other designs, e.g., using integrated optics.

\subsection{Concept}\label{sec:Concept}
The gain is provided by a multiple-emitter diode array in which
the individual gain elements are sufficiently separated from each
other to exclude mutual evanescent coupling. The array is equipped
with a high-reflection (HR) back facet and an anti-reflection (AR)
coated front. The AR coating avoids etalon effects and ensures that
the single emitters amplify light without imposing a longitudinal
mode structure. The cavity contains focusing optics, a diffraction
grating and a semiconductor saturable absorber mirror
(SESAM~\cite{Keller1996}). The external cavity confines the
feedback to each emitter to a narrow frequency range, with a
uniform spacing between the frequencies for neighboring elements.
This is achieved by arranging the diffraction grating such that
the -1st order diffraction angle is common to all emitters, such
as is described in Refs.~\cite{vGoor1983,Daneu2000}. This common
arm of the cavity is closed by the SESAM.

Each emitter receives feedback which it amplifies, providing a
continuous-wave (CW), single-frequency output per emitter. Initially, these
frequencies superimpose in the common arm with random relative phases to produce a random temporal
variation of the power. The reflectivity of the SESAM depends
on the instantaneous intensity, providing relative-phase dependent
feedback.

Mode-locking is initiated by the SESAM becoming saturated, leading
to an increased reflectivity when sufficiently many emitters
constructively interfere. This results in increased feedback to
the emitters, which drives their mutual phase locking. If the
frequencies and phases are locked, a regular train of pulses
results in the common arm. Note that in the other arm all emitters
remain CW emitters, each at its individual, single frequency. The
output pulse train can be extracted via the 0th
diffraction order.

Such a separate-gain mode-locked laser would have a number of
unusual properties. While the phases of the CW lasers are mutually
locked by the SESAM feedback, the frequencies of the individual CW
lasers are determined by the dispersion of the grating and the
physical spacing between the emitters. Note that these parameters
can be designed to select a desired frequency spacing, and thereby
repetition rate, independent of the length of the external cavity.
The average power scales with the number of gain elements and is
independent of the repetition rate, provided that the peak power
remains sufficient to saturate the SESAM. The bandwidth scales as
the number of emitters and their frequency separation and,
therefore, can be controlled by switching gain elements on and
off. A related approach with only partially separated gain was
suggested in~\cite{Bente2006} but did not separate all
longitudinal modes so that the gain elements still needed to
amplify pulses. Our approach creates pulses by synthesizing a
large number of single frequencies~\cite{Hansch1990} but without
the need for active stabilization or nonlinear conversion.

\begin{figure}[h]
\begin{center}
\includegraphics[width=1.0\textwidth]{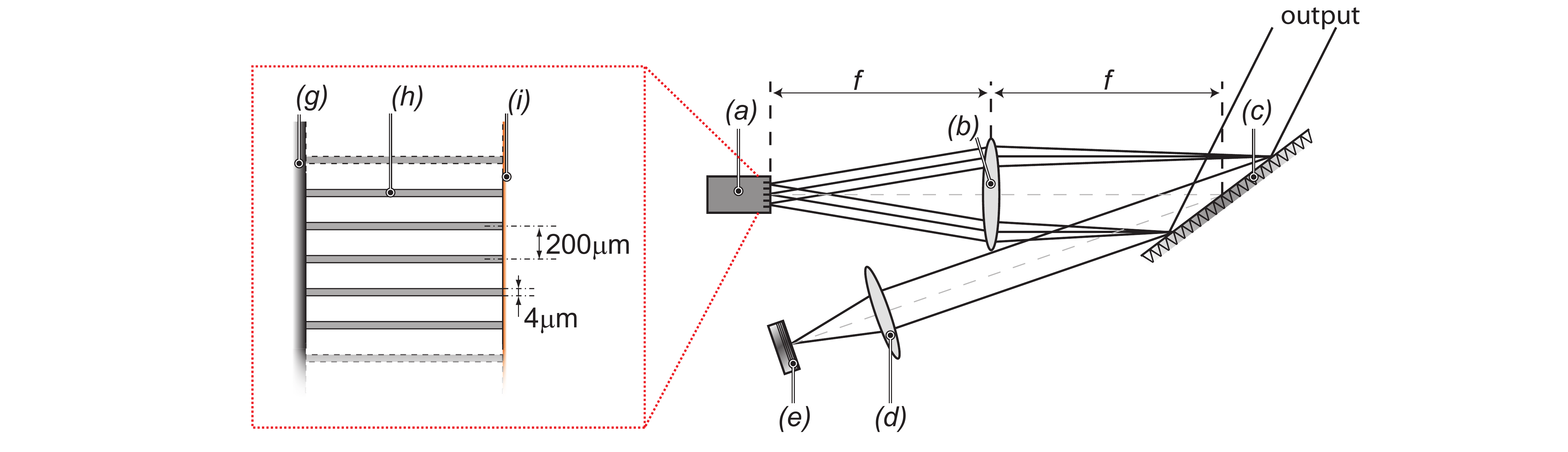}
\caption{(right) Overview. Light originates from
the AR coated diode array \emph{(a)}, and is collimated by a lens
\emph{(b)}. The light is combined by a diffraction grating
\emph{(c)}, and focused \emph{(d)} on the SESAM \emph{(e)}. The
collimation lens \emph{(b)} is placed one focal length \emph{(f)}
from the diode array, and the diffraction grating \emph{(c)} is one
focal length \emph{(f)} from the lens \emph{(b)}. (left) Magnified
view of the diode array showing several of the emitters. The gain
elements \emph{(h)} are spaced 200~$\mu$m apart and have a width of
4~$\mu$m. The array has a high-reflection on the back facet \emph{(g)} and an anti-reflection coating on the front \emph{(i)}.} \label{FigExperimentalSetup}
\end{center}
\end{figure}

In the following, we present an analytical expression that relates the
cavity design parameters to the repetition rate and pulse
duration. We also present calculations of the laser's temporal
dynamics, based on amplification and gain saturation in the slowly
varying, CW regime for each of the separate gain
elements. These calculations provide the requirements for the SESAM
to establish mode-locking. The use of a slowly varying,
CW amplifier model for each of the gain elements is
justified by the limited frequency range available to each
element, which imposes correspondingly slow dynamics. The dynamics
of the SESAM are expected to lie in the ultra-fast regime. To
account for this we employ the phenomenological model developed by
Grange~\emph{et~al.}~\cite{Keller2005}. Within this model, the SESAM is described by its recovery time,
$\tau$, modulation depth, $\Delta R$, and the saturation fluence $\Phi_{sat}$. The effect of two-photon absorption at high pulse fluences is described with another parameter, $\Phi_{TPA}$.

We expect the requirements for our system to differ from those for
conventional mode-locking where the saturation effect of the SESAM
must exceed the saturation effect of the gain, since the two
counteract each other. In conventional mode-locking, the light
flow is pulsed throughout the entire cavity. Saturation of the
SESAM lowers the losses while saturation in the gain material
lowers the gain. The competition of these effects limits the pulse
shortening. In our approach, no change in gain saturation is
experienced in the transition to pulsed operation, because each
gain element maintains a CW light field. Consequently, the SESAM
can operate over a greater range of saturation fluences. For the
practical implementation, such as in
Fig.~\ref{FigExperimentalSetup}, we expect that fast recovery times
are required because the grating cavity dispersion and
diode-element separation imply relatively high repetition rates.

\section{Model}
For the theoretical description, we used a plane wave laser model, based on a
diode array with a
large number of independent, separate single-spatial-mode gain
elements. Due to the complexity of the laser and its nonlinear
dynamics we use numerical modeling. We base the parameters in our
model on a typical single-mode diode laser array that is
commercially available (Oclaro, type~SPCxxC-980-01). The array contains 49
emitters separated by 200~$\mu$m, each with a front aperture of 4~$\mu$m
along the slow axis. The gain material is InGaAs with a spectral
gain bandwidth of 30~nm around a center wavelength of 975~nm and a
maximum output of 1~W CW per element, thus providing a substantial
output power of 49~W. The cavity, with a total length of 1~m, as
illustrated in Fig.~\ref{FigExperimentalSetup}, contains a
spherical lens ($f_{(b)}$=300~mm, diameter is 50.8~mm), a 1800~lines/mm diffraction
grating, and a semiconductor saturable absorber mirror (SESAM).
The focal length of the second lens $(d)$ is not specified.
Instead we vary the fluence on the SESAM to represent different
focusing conditions and account for losses between the emitters and the SESAM.

\begin{figure}[h]
\begin{center}
\includegraphics[width=1.0\textwidth]{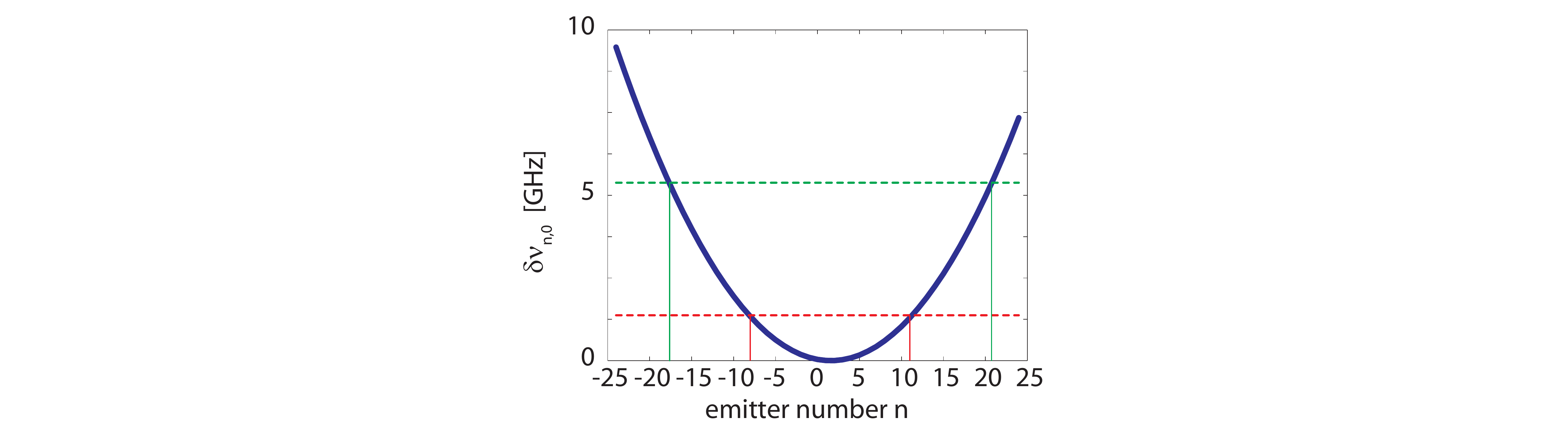}
\caption{The variation in the frequency spacing $\delta \nu_{n,0}$ in Eq.~\ref{eq:freqSeparation} versus the emitter number $n$. The red dashed line indicates the number of emitters that can be corrected assuming that the mark/space ratio of the emitters is the limiting factor. The green line indicates the number of emitters that can be corrected assuming that a 33\% modal overlap is required to frequency lock the emitter. These calculations assume that the ideal frequency spacing is 67~GHz.}
\label{FigEmitterVSdeltaOmega}
\end{center}
\end{figure}

The frequency separation between the emitted center frequencies of
the gain elements can be calculated using the standard expression for the diffraction angles at a grating. In
the small-angle approximation, the frequency separation between the
$n$th and $m$th gain element, is given by:

\begin{equation}
\delta \nu_{n,m}=\frac{c}{\Lambda_g} \bigg(\frac{1}{\alpha + \beta - \gamma \frac{nd}{f}}-\frac{1}{\alpha + \beta - \gamma \frac{md}{f}} \bigg)
    \label{eq:freqSeparation}
\end{equation}

with $n=0$ denoting the gain element located in the center of the
array (total number of element equals 2n+1), $c$ is the speed of
light, $\Lambda_g$ is the grating period, $\alpha = sin(\theta_i)$
where $\theta_i$ is the angle of diffraction of the -1st order
compared to the normal of the grating, $\beta = sin(\theta_0)$
where $\theta_0$ is the angle of the incident light at the grating
originating from the central gain element, $\gamma =
cos(\theta_0)$, $d$ is the physical distance between two adjacent
gain elements at the diode array and $f$ is the distance between
the grating and the collimating lens. Note that
Eq.~\ref{eq:freqSeparation} is independent of cavity length but is
weakly dependent of the incident light frequency, since $\theta_0$
indirectly depends on frequency. This means that an equal spacing
between the emitters does not yield an equal spacing between
adjacent frequencies, as can also be seen in
Fig.~\ref{FigEmitterVSdeltaOmega}, where the spacing varies by a
few percent. For mode-locked operation, which imposes a precisely
equidistant comb of light frequencies, this deviation is
accommodated for within the angular and spatial acceptance of the
gain elements as long as the mark/space ratio is equal to or
exceeds the deviation. In our case the mark/space ratio is 2\%
(4~$\mu$m/200~$\mu$m) and the nonlinearity in
Fig.~\ref{FigEmitterVSdeltaOmega} is 2\% over a limited range.
This means that an equally spaced frequency comb can be achieved
over a range of 19~emitters, as indicated by the red lines in
Fig.~\ref{FigEmitterVSdeltaOmega}. However, taking the numerical
aperture of the collimation lens into account, the diffraction
limited spotsize of the light fed back to the diodes has a
(1/$e^{2}$) diameter of 7.4~$\mu$m, which is larger than the
emitter aperture. This causes an overall loss but also an
insensitivity to misalignment and allows equidistant comb elements
over a larger number of emitters. For instance, if we consider
that only 33\% of the maximum modal overlap with the laser
aperture is required to ensure lasing at the correct frequency,
then an equidistant frequency comb extending over a range of
37~emitters is possible (effective mark/space of 8\%). Since diode
lasers only need a limited amount of feedback for frequency
locking, an even larger range of emitters might be frequency
locked in an equally spaced frequency comb. For optimum
performance, however, a chirped emitter spacing or suitable
correction optics should be employed. In the following
calculations, we assume that an ideal equidistant frequency comb
is possible. Eq.~\ref{eq:freqSeparation} can thus be seen as an
expression to design the repetition rate. For instance, a more
dispersive diffractive grating generates a larger frequency
spacing and, therefore, a higher repetition rate.

For the relation between pulse duration and repetition frequency,
we consider the following. The pulse repetition frequency is given
by the frequency spacing, $\delta \nu$, and the Fourier limited
pulse duration is determined by the total bandwidth, $\Delta \nu=(2n) \delta \nu$. Therefore, the ratio of the pulse duration and the pulse spacing (duty cycle) is fixed and amounts approximately to the inverse number of spaces between the gain elements, i.e., ($1/(2n)$). If a
lens system with an adjustable magnification is inserted between
the grating and the gain elements, the repetition rate and pulse
duration are simultaneously changed. To increase the pulse
duration independently of the repetition rate, a number of the
outer elements could be switched off.

For our parameters the frequency spacing is 67~GHz so that the
total bandwidth is 3.3~THz. From this we obtain a Fourier limited
pulse duration of approximately 300~fs~FWHM.

\begin{figure}[h]
\begin{center}
\includegraphics[width=1.0\textwidth]{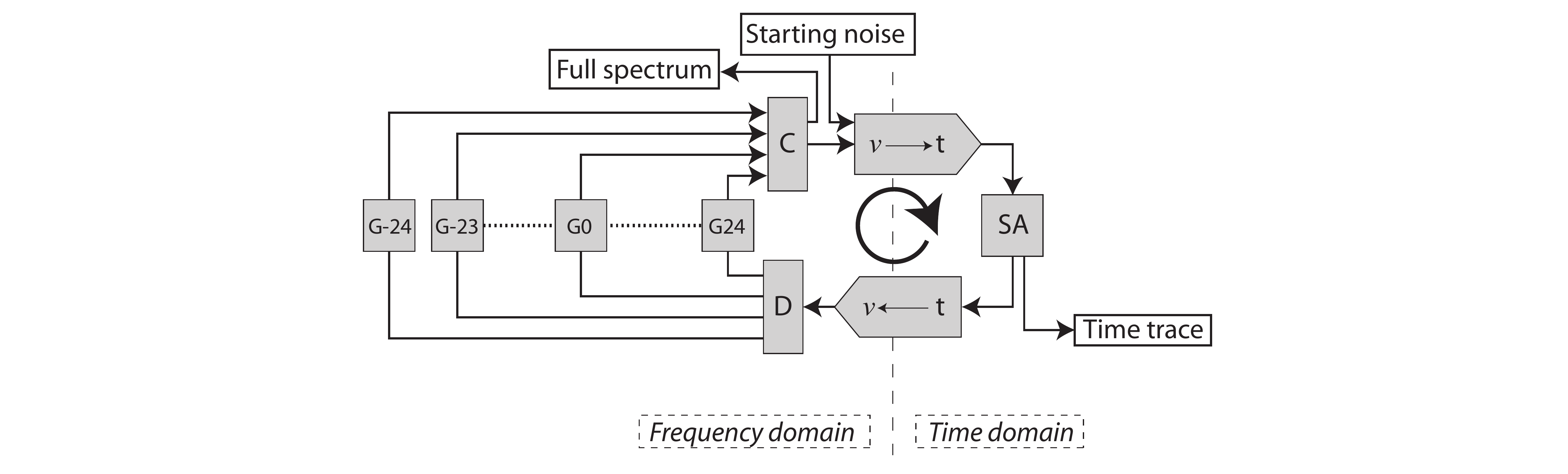}
\caption{Block diagram for the laser model. One iteration consists of light amplification in
49 independent gain elements (G-24..G24), summation of their light field amplitudes to the full spectrum (C), Fourier transformation to the time domain ($\nu \rightarrow t$), saturable
absorber described in time domain (SA) and monitoring the temporal shape of the output (time trace), inverse Fourier transformation to the frequency domain ($\nu \leftarrow t$), decomposition into separate spectral field amplitudes (D), and return to the gain elements. Random initial phases are used to model startup of mode-locking from noise.}
\label{FigBlockScheme}
\end{center}
\end{figure}

Fig.~\ref{FigBlockScheme} shows the block diagram for our
iterative calculation. We model the field envelope rather than the
oscillating field itself (slowly varying envelope approximation),
which is equivalent to shifting the frequency of the center array
element to zero. This is justified by the spectral filtering of
light by the grating inside the cavity, which reduces the light
field oscillation in the single elements to a narrow bandwidth and
corresponding slow dynamics. The model combines two essential
physical processes: the interaction of the light with the SESAM,
which is calculated in the time domain, and the amplification of
light in the gain elements, which is calculated in the frequency
domain. Switching between the frequency and time domain is done by
a discrete fast Fourier transform. Both the time and frequency
coordinates are represented by an array of $2^{14}+1=16385$
elements. The frequency spacing between each frequency array
element is chosen as 6.7~GHz, providing a bandwidth of 110~THz
over the total array length. This bandwidth provides sufficient
resolution (9.1~fs) in the time domain to describe the dynamics of
the SESAM and resolve the pulse shape. The time array spans
150~ps. The center frequencies of the gain elements are placed
67~GHz (10 array elements) apart, so that the total bandwidth
covered by the gain elements is 3.3~Thz, which corresponds to
10.5~nm at a center wavelength of 975~nm. Mode-locking will result
in 10 pulses within the time-array, all of which are processed in
one computational loop. This choice for the frequency spacing of
the gain elements corresponds to the proposed setup as discussed
earlier in this paper. However, the computational frequency
resolution is many times less than the cold cavity mode spacing,
which is $\sim$150~MHz (for a 1m total cavity length) so that more
than 400 pulses occupy the cavity simultaneously rather than 10.

\subsection{Gain model}\label{sec:GainModel}
We use a gain profile for each gain element to express two issues:
1) the possibility of oscillation on several longitudinal modes
per gain element, spaced at 150~Mhz. 2) The spectral filtering
provided by the physical size of the emitters in the diode array
and the dispersion by the grating, which we calculated to be covering about 600~MHz. As a result,
approximately five modes could oscillate per gain element. To
address the first issue, we allow for five discrete frequencies
per element, two "side modes" around each "center mode". Our
implementation with a 6.7~GHz array spacing rather than 150~Mhz
frequency spacing is chosen to reduce the calculation time, however, any
spacing can be chosen when the locking range~\cite{Kobayashi1981} exceeds the cavity
mode spacing. This is generally fulfilled for diode lasers with
external cavities where both the roundtrip loss and roundtrip gain is very high. We use an rough
estimate of 90\% loss which includes output coupling, optical
waveguide loss \cite{Byrne2009}, losses due to limited apertures,
modal mismatch of the feedback at the diode laser facets,
collimation errors \cite{Trela2009} and the non-saturable losses
of the SESAM. To address the second issue, the side modes
are given a lower relative gain than the center mode. The spectral
gain profile for each gain element was chosen as \mbox{[0.8 0.9
1.0 0.9 0.8]}, which we estimated as the frequency dependent
decrease in spatial overlap of the center mode and the first and
second side modes.

For the overall shape of the gain spectrum, from the -nth to the
nth center modes, i.e., for the total bandwidth of the modeled laser,
we assume a flat profile. This expresses that this bandwidth is
substantially smaller than that of the InGaAs gain \cite{Eng1990}.
Outside the emitters no gain is present.

The spatial separation of the gain per element resembles a perfect
inhomogeneous broadening. This implies an insensitivity to
residual etalon effects which are known to disturb mode-locking in the
conventional approach.

Since we are interested in the dynamics rather than the power, the
total amount of power inside the laser is normalized to 49~W (1~W
per gain element) without reference to the transverse size. This
is done by normalizing the saturation intensity of the gain to the maximum signal intensity inside the gain element (based on a power of 1~W). This yields
\mbox{$I_{sat}=1/(g_0-1)$}, which normalizes the extractable power
\cite{BookSiegman1986} and fixes the relationship between the
small signal gain and the saturation intensity. The remaining
degree of freedom is the value of the small signal gain, $g_0$,
which we choose to be a typical value of 100 over the full length of the gain. Roundtrip losses are
set to a typical value of 90\% and are taken to include all losses
as described before. These losses are applied to all the light
that is fed back into the individual gain elements before
calculating the gain. For the gain of each individual element we assume homogeneous broadening which implies that, upon amplification, the center mode and the side modes compete for
gain. This is approximated as \cite{BookSiegman1986}:
\begin{equation}
g(I_{tot})=\frac{g_0}{1+I_{tot}/I_{sat}}
\label{eq:gain}
\end{equation}
where $I_{tot}$ is the sum of intensities of the center and side
modes. The gain in each iteration is calculated using
Eq.~\ref{eq:gain} rather than using an integration over the full
volume of the gain element. In the initial iterations this approximation does not strictly conserve
energy per roundtrip. But, in steady-state, the saturation reduces the roundtrip gain to a value of 1 for a per-emitter power of 1~W, conserving power over
multiple round trips and therefore the energy. The amplified
spectrum is transformed to the time domain using a fast
Fourier transform.

\subsection{Temporal dynamics}
The reflection of the SESAM is described in the time domain. We
follow a phenomenological approach developed by Grange~\emph{et~al.}~\cite{Keller2005}, which gives the reflectivity of a SESAM as
\begin{equation}
R(\Phi_P)=R_{ns}-(1-e^{-\Phi_P/\Phi_{sat}})\frac{\Delta R}{\Phi_P/\Phi_{sat}}-\frac{\Phi_P}{\Phi_{TPA}}.
\label{EqTPA}
\end{equation}
Here $\Phi_P$ is the pulse fluence (for a pulse much shorter than the recovery time
$\tau$), $\Delta R$ is the modulation depth of the reflectivity,
and $R_{ns}$ incorporates the non-saturable losses. Two-photon
absorption is described using a two-photon fluence $\Phi_{TPA}$,
where $\Phi_{TPA}=S_0^2 \Phi_{sat}/\Delta R$ \cite{Keller2005} and
$S_0=10$, for a typical commercially available SESAM
\cite{WebsiteBatop}. Two-photon absorption accounts for a decrease in reflectivity with increasing pulse fluence beyond a certain point ($S_0$). As the temporal and spectrally independent
cavity losses are combined into a single loss term, which is taken
into account in the gain section of the model, we set $R_{ns} =1$.

In an experimental setting, the fluence level on the SESAM can be
varied continuously by changing the cavity mode diameter at the
SESAM or increasing the pump current through the gain elements. In our
calculations the cavity mode diameter is not specified and the
total power is normalized. The pulse energy is fixed and the pulse
fluence only has relevance with respect to the saturation fluence
of the SESAM. We define a parameter $S$ as the ratio of the pulse
fluence for a Fourier limited pulse over the saturation fluence of
the SESAM: $S=\Phi_{opt}/\Phi_{sat}$. This ratio defines the
effective size of the beam on the SESAM and allows for a pulse
energy to be converted to a pulse fluence.

The $\Phi$'s in Eq.~\ref{EqTPA} are pulse fluences for pulses much
shorter than the recovery time and therefore have no explicit time
dependence. The fluences correspond to the saturation level in the
SESAM in a time interval too short to experience the exponential
decay of the saturation over the recovery time $\tau$. To derive
this "short-pulse fluence" $\Phi_P$ from the temporal intensity we
use Eq.~\ref{eq:SESAMAbs}, in which the integration of the
intensity is weighted by the exponential decay of the SESAM, i.e., the recovery time.
Please note that this pulse fluence can only be applied when used in
combination with the SESAM's recovery time and does not describe the
pulse energy directly.
\begin{equation}
    \Phi_{P} = \int I(t)\exp(-t/\tau)dt
    \label{eq:SESAMAbs}
\end{equation}
In this expression, $\tau$, represents the recovery time of the
SESAM. The integration is performed as a running sum and the
resulting value for $\Phi_{P}$ is inserted into Eq.~\ref{EqTPA} to
obtain the time-dependent intensity reflectivity, $R(\Phi_P)$. The
time-domain field amplitude is multiplied by the square root of
$R(\Phi_P)$. This concludes the time-domain portion of the
calculation iteration and the time-domain field amplitude is
transformed by discrete fast Fourier transform to the frequency
domain to begin a new iteration.

\subsection{Initial conditions and evaluation}
The calculations are started with a power of 1~W per gain element
divided over the main mode and the side modes in the shape of the spectral
gain profile mentioned before. To model the influence of spontaneous emission on the onset of mode-locking, the phases of the modes are
assigned new random values at the beginning of each simulation.
For all the data reported here, the simulations were terminated at
500 iterations because it was observed that a steady-state was
generally reached within 500 iterations. To observe the impact of
the initial phases to mode-locking, we ran each simulation multiple
times with the same SESAM parameters.

Simulations were performed for a wide range of the SESAM parameters $S$, $\tau$,
and $\Delta R$. To evaluate the numerous time traces, we defined a criterion for successful mode-locking by comparing each calculated time trace with that of a Fourier limited pulse train that possesses the maximum possible peak power. Successful mode-locking was defined as the occurrence of any peak power that exceeds half of this maximum possible peak power. With this criterion we obtain the probability of mode-locking by determining how many times, out of a given number of simulations with identical parameters, successful mode-locking was observed. The results showed that ten runs are sufficient to identify a parameter range where mode-locking occurs, which we call the mode-locking regime, with a distinguishable boundary to parameters with which no mode-locking occurs. So we have now defined the probability of mode-locking as the number of times mode-locking was observed, divided by the number of simulations per parameter set (which we chose to be ten).

\section{Results}
Based on Eq.~\ref{EqTPA}, we expect to see mode-locking for values
of $S$ above $S=1$, where the pulse fluence equals the
SESAM's saturation fluence. At $S=10$ we expect mode-locking over
the largest range of SESAM parameters, $\Delta R$ and $\tau$. For higher
values of $S$, however, two-photon absorption is expected to decrease the
mode-locking regime. We show results for $S$: 0.1, 1, 3, 5, 10, 20,
and 50. The values for the SESAM's recovery time and modulation
depth are chosen similar to what current, commercially available SESAMs provide
\cite{WebsiteBatop}: $\Delta R$ is chosen between 5\% and 40\%, in
steps of 5\%, and $\tau$ is chosen between 0.5~ps and 10~ps.

\begin{figure}[ht]
\begin{center}
\includegraphics[width=1.0\textwidth]{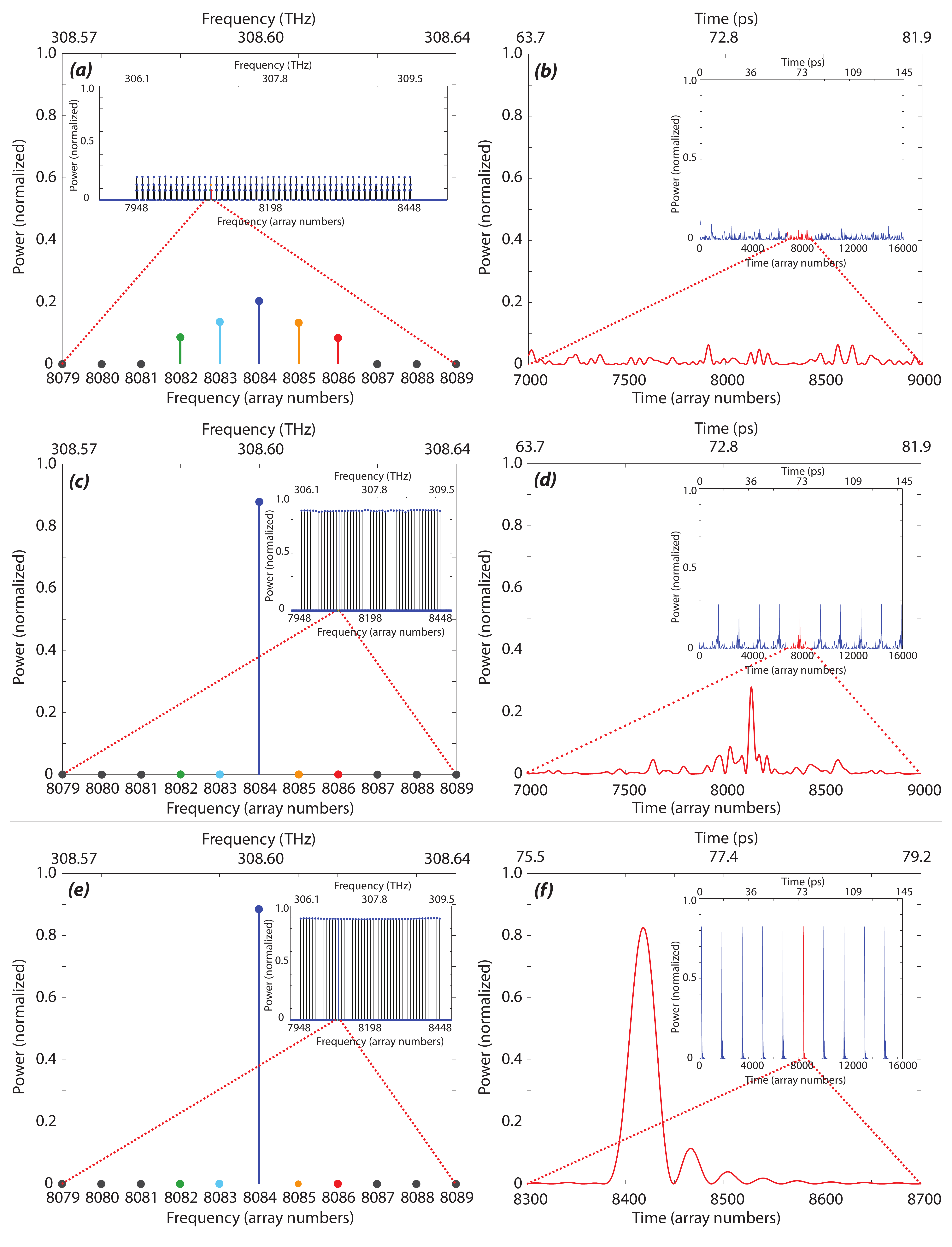}
\caption{Calculation results after 1 \emph{(a,b)}, 100 \emph{(c,d)} and 500
\emph{(e,f)} iterations, for $\Delta R = 40\%$, $\tau = 3~ps$ and $S=3$.
The power spectrum is shown in the left column with the
corresponding time-trace on the right. All power spectra are
normalized to the maximum power for 1~W per emitter. The
time-traces are normalized to the Fourier limited maximum peak power.
The laser starts with multiple frequencies per gain element and a
random phase spectrum. After 500 iterations the laser is
mode-locked} \label{FigPulsesAndSpectra}
\end{center}
\end{figure}

To show how the pulse train and spectrum evolve, three different
iterations are shown in Fig.~\ref{FigPulsesAndSpectra}. For these calculations the SESAM
parameters were chosen at the edge of the mode-locking regime
($\Delta R = 40~\%$, $\tau = 3~ps$ and $S=3$), to have a clear
evolution of the pulses over a large range of the 500 iterations.
Fig.~\ref{FigPulsesAndSpectra} \emph{(a)} and \emph{(b)} show the
spectral power and corresponding time-trace, after one
iteration, \emph{(c)} and \emph{(d)} after 100 iterations, and
\emph{(e)} and \emph{(f)} after 500 iterations. The insets in
\emph{(a)}, \emph{(c)} and \emph{(e)} show the spectrum including
all side-modes over the range of the gain. \emph{(a)}, \emph{(c)}
and \emph{(e)} show a zoom-in of the spectrum to ten array numbers
(array numbers 8079-8089, the distance between two neighboring
center modes). The dots (and vertical lines) show the calculated power of the light
at these frequencies. The insets in
\emph{(b)}, \emph{(d)} and \emph{(f)} show the full time-trace of
our calculations. \emph{(b)} and \emph{(d)} show a zoom-in over
2000 array numbers. \emph{(f)} shows a zoom-in over only 400 array
numbers to resolve the pulse shape at the end of the calculation.

In Fig.~\ref{FigPulsesAndSpectra}\emph{(a)}, after one iteration, it can be seen that all
side-modes are still present at the spectrum, however their
power relative to the center frequency is lower due to gain
competition and a lower relative gain. After 100 iterations (Fig.~\ref{FigPulsesAndSpectra}\emph{(c)}), no
significant power is observed for the side-modes anymore and
the available power is present in the center frequencies. Furthermore, one sees that the power at the different gain elements is not precisely equal anymore (the envelope over the full laser spectrum is not entirely smooth; the largest difference in height between the adjacent frequencies is 1.6\% the average it is 0.3\%). This can be addressed to the phase dependent exchange of energy between the modes caused by sideband generation in the SESAM. After 500 iterations (Fig.~\ref{FigPulsesAndSpectra}\emph{(e)}), the
envelope is smooth again, which is an indication for mode-locking
\cite{Garner1992} (here the largest difference in height between the adjacent frequencies is 0.23\% the average it is 0.05\%, showing that the spectrum in Fig.~\ref{FigPulsesAndSpectra}\emph{(e)} is significantly smoother than the spectrum in Fig.~\ref{FigPulsesAndSpectra}\emph{(c)}). The spectrum has an almost square-shaped envelope, which is expected
because the spatial separation of the gain for the 49 light frequencies acts as an inhomogeneous broadening of the gain. In the time-trace after one
iteration (Fig.~\ref{FigPulsesAndSpectra}\emph{(b)}), only a randomly fluctuating power is observed,  indicating that all the frequencies have a random phase. After
100 iterations (Fig.~\ref{FigPulsesAndSpectra}\emph{(d)}), the initial stages of a pulse train can be
observed, with irregularly shaped pulses, but at a regular interval. Since not
all phases in the spectrum are locked, the peak power is still
rather low. After 500 iterations (Fig.~\ref{FigPulsesAndSpectra}\emph{(f)}), short, almost Fourier-limited, pulses are observed, with a FWHM pulse duration of 273~fs (i.e., 30 array
numbers).

The remaining deviation from the Fourier limit is mainly due to the asymmetric shape of the pulses. The asymmetry can be explained by the fluence-dependent reflection of the SESAM that preferentially absorbs the front of the pulses. This preferential absorption of the front also induces a slow shift (6.5~fs per iteration) of the pulse as the iterations progress. The side pulses, seen only behind the main pulse because of the named asymmetry, are due to the shape of the laser's overall spectrum. The square-shaped spectrum, in the Fourier limit, would provide sinc$^2$-shaped pulses, with side pulses.

The most important conclusion from the results in
Fig.~\ref{FigPulsesAndSpectra} is that our novel approach to
mode-locking appears promising for an experimental demonstration
since the calculations are based on realistic parameters, i.e.,
taken from the specifications of commercially available
components, particularly with regard to the parameters of the
SESAM.

\begin{figure}[ht]
\begin{center}
\includegraphics[width=1.0\textwidth]{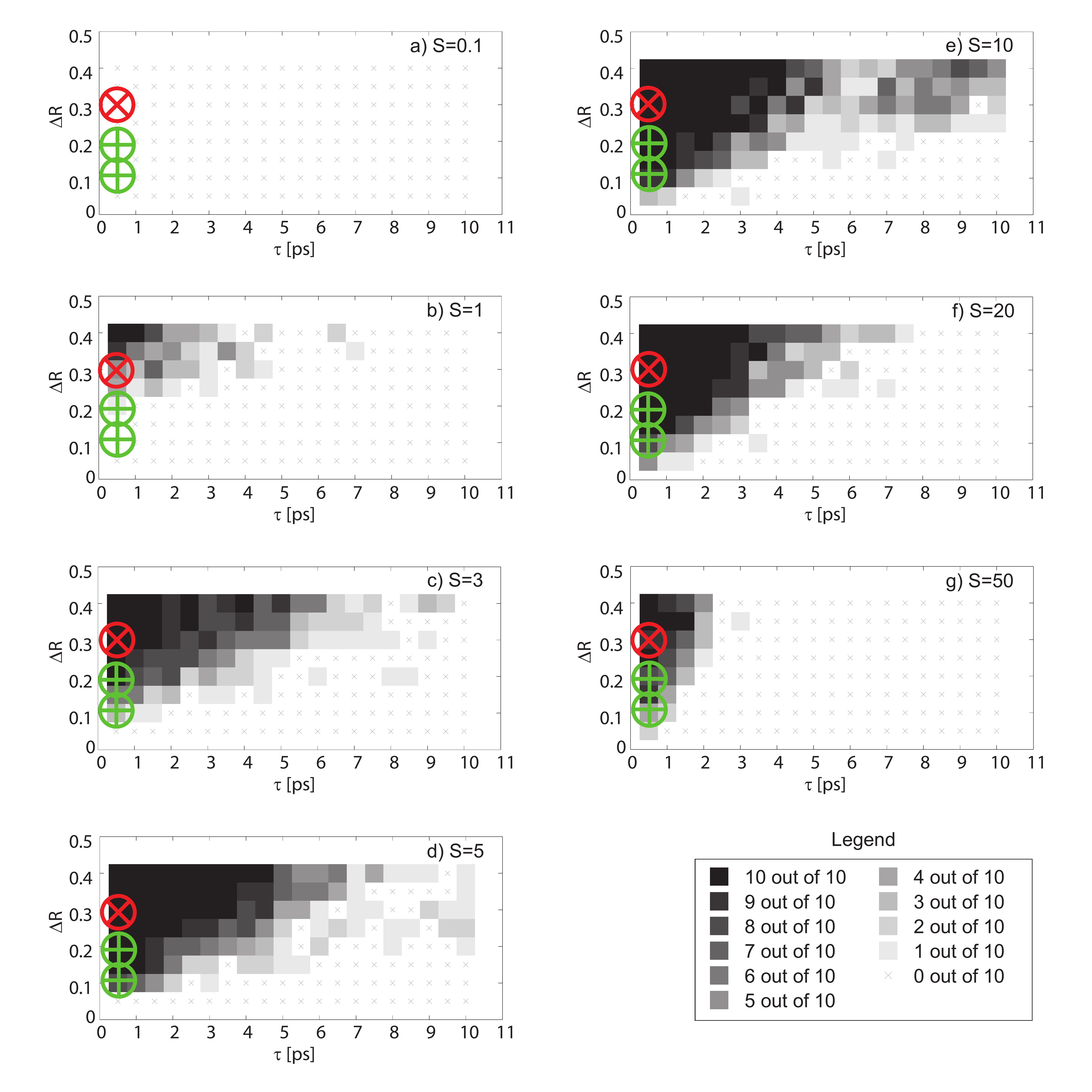}
\caption{Probability of mode-locking as a function of $\Delta R$
and $\tau$ for \emph{(a)} $S=0.1$, \emph{(b)} $S=1$, \emph{(c)} $S=3$, \emph{(d)} $S=5$, \emph{(e)}
$S=10$, \emph{(f)} $S=20$ and \emph{(g)} $S=50$. Darker shades represent areas
where mode-locking was observed more often. $\otimes$ indicates
the most suitable SESAM for our laser that is also commercially available at this time, i.e., the SESAM with the highest $\Delta R$ and the
lowest $\tau$. For comparison, the second and third best available
SESAMs are indicated by the $\oplus$ \cite{WebsiteBatop}.}
\label{FigResultsAvsT}
\end{center}
\end{figure}

We investigated the probability of mode-locking as defined above for \mbox{$0.1 < S < 50$}, \mbox{5\% $< \Delta R <$ 40\%} and \mbox{0.2 ~ps $<\tau <$ 10~ps}.
Fig~\ref{FigResultsAvsT} shows the observed mode-locking
probability in graphs with pair wise variation of $\tau$ and
$\Delta R$ for various values of $S$. Black indicates a
mode-locking probability of unity and mode-locking was observed
less often for lighter shades of gray. The small crosses in the
figure represent the calculations in which mode-locking was not
observed. We found that generally for low values of $S$ (corresponding to weaker focusing on the SESAM or a lower saturation fluence), mode-locking is only observed for fast absorbers (small~$\tau$) and absorbers with a high modulation depth (large~$\Delta R$). This is expected, because, in Eq.~\ref{EqTPA} for small values of $S$, the SESAM's reflectivity shows a modulation of less than $\Delta R$, making the loss differential too small to induce mode-locking.

As $S$ increases with sharper focusing on the SESAM, so does the parameter range for which
mode-locking successfully occurs. This is the case because, for larger $S$,
the modulation due to saturation increases, which relaxes the requirements for the modulation depth and response time. However, it can be seen in Fig.~\ref{FigResultsAvsT}\emph{(f)} and \emph{(g)} that a further increased fluence in the SESAM ($S=20$ and $S=50$, respectively) narrows the mode-locking regime for $\Delta R$ and $\tau$ again. This can be understood, because in Eq.~\ref{EqTPA} for values of $S>10$, two-photon absorption becomes significant enough to reduce the modulation of the reflectivity. In general, selecting a larger~$\Delta R$
means that the same reflectivity change, and thus the same effect on the laser dynamics, is reached for a lower fluence, broadening the operating regime for which mode-locking is
observed. Similarly, a short~$\tau$ ensures a stronger
differential gain for the pulsed state which improves the
mode-locking.

Current SESAMs are optimized for the standard approach in
mode-locking, i.e., for rather low (tens~of~MHz) repetition rates, so that longer recovery times, in the order of a several~ns, are sufficient to induce mode-locking. In our
approach, the pulse repetition frequencies can be much higher, such as 67~GHz (pulse spacing 15~ps) for the laser described here, and significantly
shorter recovery times are required. That these can indeed be obtained, though at the expense of a somewhat reduced modulation depth, can be seen in Fig.~\ref{FigResultsAvsT}. There, the best commercially available SESAMs we could identify at this moment for our laser \cite{WebsiteBatop} are indicated as symbols ($\otimes$ and $\oplus$), with a recovery time of about $\tau>0.5~ps$ and with modulation depths, $\Delta R$, between $15$~and~$30\%$. The calculations show that such values should suffice for mode-locking if a sufficiently high fluence of the incident pulses, about $S=10$, can be realized in spite of the high repetition rate. If we
assume a mode area at the SESAM of $50~\mu m^2$ and a typical
SESAM value for \mbox{$\Phi_{sat}$ of 60~$\mu J/cm^2$}, it would be
possible to achieve $S=10$ with a total power of 20~W in the
common arm. This power is a value which is well below the specifications of the diode arrays that we consider here. This estimate, thus, indicates that our novel mode-locking scheme is just feasible with currently available components.

\section{Conclusion and outlook}
We have presented a novel approach to mode-locking, based on
spatially separated gain elements and passive mode-locking. We
have numerically modeled the basic spectral and temporal properties at the example
of a multiple element diode laser (diode laser array) with an
external cavity, containing a diffraction grating and a
semiconductor saturable absorber mirror (SESAM). Our modeling
reveals the influence of the physical properties of the SESAM on
the feasibility of mode-locking. The ongoing development of faster
SESAMs with a higher modulation depth \cite{Keller2006} makes it
realistic to assume that mode-locking over a wide range of average
powers and repetition rates will be achievable. Our approach
to mode-locking lifts current limitations on the combination of
repetition rate and output power. Although we have used our model
to analyze mode-locking of a specific laser, a diode laser array
with 49~W average output power, 67~GHz repetition rate and 300~fs
pulse duration, the model can also be applied to investigate
separate gain mode-locking requirements for other types of lasers or at different
combinations of output power, repetition rate and pulse duration.
\end{document}